\documentclass[twocolumn,aps,prl]{revtex4-1}

\usepackage{amsmath}
\usepackage{graphicx}
\usepackage{esint}
\usepackage{setspace}
\usepackage{epsfig}
\usepackage{color}

\def\beq{\begin{equation}}
\def\eeq{\end{equation}}
\def\beqa{\begin{eqnarray}}
\def\eeqa{\end{eqnarray}}

\def\cH{{\mathcal H}}

\def\Tr{\mathrm{Tr}}
\def\r{{\bf r}}

\newcommand{\bra}[1]{| #1 \rangle}
\newcommand{\ket}[1]{\langle #1 |}

\begin{document}  

\author{Elinor Zerah-Harush$^1$ and Yonatan Dubi$^1,^2$}
\email{jdubi@bgu.ac.il}
\affiliation{$^1$Department of Chemistry and $^2$Ilse-Katz Institute for Nanoscale Science and Technology, Ben-Gurion University of the Negev, Beer-Sheva 84105, Israel}

\title{Enhanced Thermoelectric Performance in Hybrid Nanoparticle--Single Molecule Junctions}
\date{\today}
\begin{abstract}
It was recently suggested that molecular junctions would be excellent elements for efficient and high-power thermoelectric energy conversion devices. However, experimental measurements of thermoelectric conversion 
in molecular junctions have indicated rather poor efficiency, raising the question of whether it is indeed possible to design a setup for molecular junctions that will exhibit enhanced thermoelectric 
performance. Here we suggest that hybrid single-molecule nanoparticle junctions can serve as efficient thermoelectric converters. The introduction of a semiconducting nanoparticle introduces new 
tuning capabilities, which are absent in conventional metal-molecule-metal junctions. Using a generic model for the molecule and nanoparticle with realistic parameters, we demonstrate that the thermopower 
can be of the order of hundreds of microvolts per degree Kelvin, and that the thermoelectric figure of merit can reach values close to one, an improvement of four orders of magnitude improvement over existing 
measurements. This favorable performance persists over a wide range of experimentally relevant parameters, and is robust against disorder (in the form of surface-attached molecules) and against 
electron decoherence at the nanoparticle molecule interface.\end{abstract}
\maketitle   


\section{Introduction} 
In recent years, the study of single-molecule junctions -- the ultimate limit of electronic nano-technology -- has progressed well beyond their role as functional elements in electronic devices
 \cite{Aviram1974}, and today additional functionalities, from opto-electronics and spintronics, through phononics, to thermoelectricity \cite{Aradhya2013,Tsutsui2012,Dubi2011} are under investigation. The potential applicability of single-molecule junctions in thermoelectricity -- the conversion of heat into electric power -- is of particular interest, since thermoelectricity may turn out to be an important element in addressing global  
energy issues \cite{Bell2008,Wang2014b}. The clear advantages of thermoelectric energy conversion, such as the "green" nature of the energy conversion process, its applicability for waste heat harvesting, and the easy device maintenance due to the absence of moving 
parts, would lead us to think that thermoelectric devices should already be a substantial part of the energy market, yet this is not the case. The reason lies in 
the simple fact that current thermoelectric devices are not efficient enough, making competition with traditional (large scale) energy conversion systems virtually impossible. 

Due to their versatility, low dimensionality, and low thermal conductivity, molecular junctions offer a possible route for enhancing thermoelectric performance \cite{Mahan1996}. Indeed, 
various theoretical studies have suggested that molecular junctions could be a key component of efficient and high-power thermoelectric devices \cite{Murphy2008,Bergfield2009,Bergfield2010,Finch2009,Karlstrom2011,Nozaki2010,Stadler2011,Nakamura2013,Vacek2015}. 
Experimental demonstrations \cite{Malen2009,Malen2009a,Malen2010,Reddy2007,Widawsky2012,Widawsky2013,Chang2014,Kim2014,Balachandran2012,Lee2014}, however, have indicated just the opposite, as can be deduced by examining the two central parameters of thermoelectric conversion, the thermopower (or the Seebeck 
coefficient) and the thermoelectric figure of merit (FOM). 

The thermopower $S$ measures the voltage generated per unit temperature difference (in the linear response) \cite{DiVentra2008,Scheer2010,Malen2010,Dubi2011}. Values of $S\sim 10^2 - 10^3$ $\mu$V/K are typical for standard semi-conductor based thermoelectrics, yet molecular junctions exhibit small values of $S$, typically $S\sim 5-50$ $\mu$V/K. 
The FOM, namely $ZT$, defined as $ZT=\frac{G S^2}{\kappa /T}$, where $G$ is the conductance of the junction, $\kappa=\kappa_e+\kappa_{phn}$ is the total thermal 
conductance, which includes both electronic (e) and phononic (phn) contributions, and $T$ is the temperature. $ZT$ is directly related to the device efficiency \cite{Goupil2011}, and $ZT\rightarrow \infty$ corresponds to the Carnot efficiency (thus, theoretically, there is no upper bound on $ZT$). It is commonly held that, from the efficiency perspective, a $ZT$ of $\sim 4$ is required for thermoelectric conversion to be competitive \cite{Vining2009}. However, typical $ZT$ values obtained from measurements in molecular junctions are $ZT\sim 10^{-3}-10^{-5}$.  Thus, there seems to be a discrepancy between theoretical and computational studies of thermoelectric conversion in molecular junctions, which in many cases \cite{Murphy2008,Bergfield2009,Bergfield2010,Finch2009,Karlstrom2011,Nozaki2010,Stadler2011} predict $S\sim 10^2-10^3$ $\mu$V/K and FOM of 
$ZT \gg 1$, and the experimental evaluation of the two measures. 

The origin of this discrepancy seems to stem from two factors. First, one has to be careful to take the phonon contribution to the thermal 
conductance into account; a typical and realistic value for the phonon thermal conductance in molecular junctions is $\kappa_{phn}=10-100$ pW/K \cite{Malen2010,Ong2014,Segal2003,Wang2007,Nozaki2010}. Second, many  
calculations show enhanced thermoelectric performance based on tuning of the molecular orbitals or, equivalently, the Fermi level of the electrodes. However, in reality this tuning is very difficult to 
achieve \cite{Ballmann2012,Capozzi2014,Kim2014,Prins2011}, and the only "tuning parameter" available for molecular junctions is the choice of the molecular moiety. It is thus a central challenge to design  
molecular junctions that are both tunable in some way and show favorable thermoelectric performance with realistic parameters. 

In this paper, we present a new setup for molecular junctions that has the potential to achieve this goal. Our setup is based on a hybrid single-molecule semiconducting nanoparticle  (SC-NP) 
structure in which a molecule is connected on one side to a metallic electrode (as in conventional molecular junctions), and on the other side connected to the second electrode through a SC-NP placed on the electrode, as is 
schematically depicted in Fig.~\ref{fig1}(a). In this setup, the molecular junction can be tuned via tuning the electronic properties 
of the nanoparticle, namely, by choosing a suitable material and by controlling the size and shape of the nanoparticle (for reviews see, e.g., \cite{Klimov2003,Banin2003,Talapin2010}). Fabrication of hybrid NP-single moelcule 
junctions have already been demonstrated with Au NPs (see, e.g., \cite{Liao2006,Jafri2010,Huang2006}), making our suggested system experimentally feasible.

Using a generic model for the SC-NP, we show that this junction can reach values of $S\sim 100-400$ $\mu$V/K and $ZT\sim 1$ for a broad range of realistic parameters. The origin of the enhanced thermoelectric 
performance can be traced to the interplay between the local transport properties of the molecule and the gapped density of states (DOS) of the nanoparticle \cite{Nozaki2010}. Thermoelectricity typically requires a large particle-hole asymmetry (reflected in the fact that, at low temperatures, the thermopower is proportional to the derivative of the transmission function 
\cite{Scheer2010,Malen2010,Dubi2011}). This asymmetry is enhanced in this junction due to the presence of the semiconducting gap in the SC-NP \cite{Nozaki2010}, an effect that is rectified due to the 
finite size of the nanoparticle. We show that the optimal parameters for thermoelectric conversion depend on the geometry of the nanoparticle and the contact geometry between the nanoparticle and the 
molecule. Further, we demonstrate that the favorable 
thermoelectric performance is robust against disorder (in the form of surface dangling molecules) and dephasing, and finally, we discuss the temperature dependence of the FOM, which exhibits a maximum at $T \sim 450$ K. 

\section{Model and Calculation}

The transport and thermoelectric properties of the hybrid molecule-nanoparticle junction are calculated by using the non-equilibrium Green's function approach, which has become the standard tool for such calculations 
\cite{DiVentra2008,Scheer2010}. The junction (graphically depicted in Fig.~\ref{fig1}(a)) is described using the Hamiltonian:
\beq
\cH=\cH_B+\cH_{NP}+\cH_M+\cH_T+\cH_{B-NP}+\cH_{NP-M}+\cH_{M-T}~~, \label{H1} \eeq
that includes the bottom electrode (B), the nanoparticle (NP), the molecule (M), the top electrode (T), and the coupling between the bottom electrode and nanoparticle (B-NP), between the NP and the molecule (NP-M) and between the 
molecule and the top electrode (M-T). Since electron spin does not play a role in the mechanisms we describe here for enhancement of thermoelectricity, we treat spinless electrons. 

To describe the SC-NP, we use a generic tight-binding model for a semiconductor \cite{Fainberg2012,Reuter2009,Delerue2004,Schmid2011,Diaz2004,Hill1994} of the form 
\beq
\cH_{NP}=\sum_{\r} (\epsilon_{c}+\frac{\Delta}{2} (-1)^{P_\r}) c^{\dagger}_\r c_\r - t\sum_{\langle \r \r' \rangle}c^{\dagger}_\r c_{\r'} ~~,\label{H-NP} \eeq
where the summation is taken over the atom positions $\r$ (assumed to form a cubic lattice); $c^{\dagger}_\r (c_\r)$ creates (annihilates) an electron at position $\r$; $t$ is the nearest-neighbor 
hopping matrix element; $\epsilon_c$ is the position of the band-gap center; and $\Delta$ is the semiconductor band-gap. The function $P_\r=x+y+z$ gives 
a modulation of the on-site energy $\epsilon_c \pm \frac{\Delta}{2} $ between neighboring atoms, generating a gapped band at the thermodynamic limit \cite{Reuter2009}. We point that although {\sl ab 
initio} methods for calculating properties of NPs are emerging \cite{Li2007,Azpiroz2014}, these are still not developed for transport calculations. Furthermore, since we are aiming at presenting 
general properties of hybrid junctions, our tight-binding calculation is generic and not limited to a specific system.

The molecule is described as a single orbital, which can correspond to either the HOMO or LUMO, depending on the position of the orbital energy \cite{Scheer2010}. The molecular Hamiltonian is simply 
\beq
\cH_M=\epsilon_0 d^\dagger d ~~, \eeq
where $d\dagger (d)$ creates (annihilates) an electron at the molecular orbital. In addition to the calculations described below, we performed calculations that also include a Coulomb 
interaction term (for the molecule) and spin-full fermions (by using the equations-of-motion method \cite{Meir1991,Souza2007}) but found no quantitative change in the main results, and therefore we keep 
the description here as simple as possible and treat electrons as non-interacting particles. The coupling between the molecule and the nanoparticle is described by the Hamiltonian: 
\beq 
\cH_{NP-M}=-t_1 c^{\dagger}_{\r_M} d+h.c.~~,\label{H2} \eeq where $\r_M$ is the position of the atom in the nanoparticle that is in contact with the molecule, and $t_1$ is the hopping matrix element related to the overlap integral between the molecular orbital and the atomic level at $\r_M$.

The metallic top and bottom electrodes are assumed to be non-interacting metals \cite{Meir1992} with the Hamiltonian $\cH_{X}=\sum_k \epsilon_k c^{\dagger}_{k,X}c_{k,X},~~X=T,B$ 
\cite{Meir1992,DiVentra2008,Scheer2010}. The coupling between the molecule and the top electrode (representing, e.g., the tip of a scanning tunneling microscope) is represented by the Hamiltonian term 
$\cH_{M-T}=\sum_{k,T} V_{k,T} c^\dagger_{k,T}d +h.c.$. Similarly, the contact between the bottom electrode and the nanoparticle is described by $\cH_{B-MP}=\sum_{\r \in B} V_{k,B} c^\dagger_{k,B} c_{\r}+h.c.$. 
To proceed with the transport calculations, the metallic electrodes are treated within a wide-band approximation \cite{Verzijl2013} (valid since our model is for non-interacting electrodes), where the 
top electrode is defined by a (retarded and advanced) self-energy term 
$\Sigma^{r,a}_{T}=\mp i \frac{\Gamma_{T}}{2} \bra{M}\ket{M}$,  in which $\bra{M}$ is the single-particle molecular orbital, and $\Gamma_{T}$ is the electrode-induced level broadening. Similarly, the 
bottom electrode is defined via the self-energy term $\Sigma^{r,a}_{B}= \mp i \frac{\Gamma_{B}}{2} \sum_{\r \in B} \bra{\r}\ket{\r}$, where  $\Gamma_{B}$ is the broadening due to the bottom electrode 
and the summation is taken over all the atom positions in the nanoparticle that are in contact with the bottom electrode. We point that the model described above (and specifically the wide band approximation) implies electron-phonon-induced thermalization in the electrodes, but no electron-phonon interaction in the junction at this stage. All the relevant energies- the Fermi level of the electrodes, NP valence and conduction bands, and molecular orbitals -  are schematically also shown in Fig.~\ref{fig1}(a). 

Once the Hamiltonian and the self-energies are defined, the calculation proceeds via the non-equilibrium Green's function approach, which is reduced to the Landauer formalism for non-interacting electrons. \cite{Datta1997,DiVentra2008,Peskin2010}. The Green's functions are determined via $G^{r,a}=\left( E-\cH+\Sigma^{r,a} \right) ^{-1}$, where $\Sigma^{r,a}=\Sigma^{r,a}_T
+\Sigma^{r,a}_B$. The transmission function is given by $T(E)=\Tr \left( \Sigma^r_T G^r \Sigma^a_B G^a \right) $, and the transport coefficients, namely, the conductance $G$, the thermopower $S$, 
and the thermal conductance $\kappa$, are determined within the Landauer formalism as $G=e^2 L_0,~S=L_1/(e T L_0,~\kappa=\left( L_2-\frac{L_1^2}{L_0}\right)/T$, where $T$ is the temperature (room temperature, unless otherwise stated) and $L_n=-\frac{1}{h}\int dE T(E)(E-\mu)^n \frac{\partial f}{\partial E}$ are the Landauer integrals, with $h$ being Planck's constant, $\mu$ the chemical potential of the electrodes, and $f(E)$ the Fermi-Dirac distributions. The thermoelectric FOM $ZT$ is given by $ZT=\frac{G S^2}{\kappa/T}$.

\section{Results} 

\begin{figure}[t]
 \vskip 0.0truecm
 \includegraphics[width=8.5truecm]{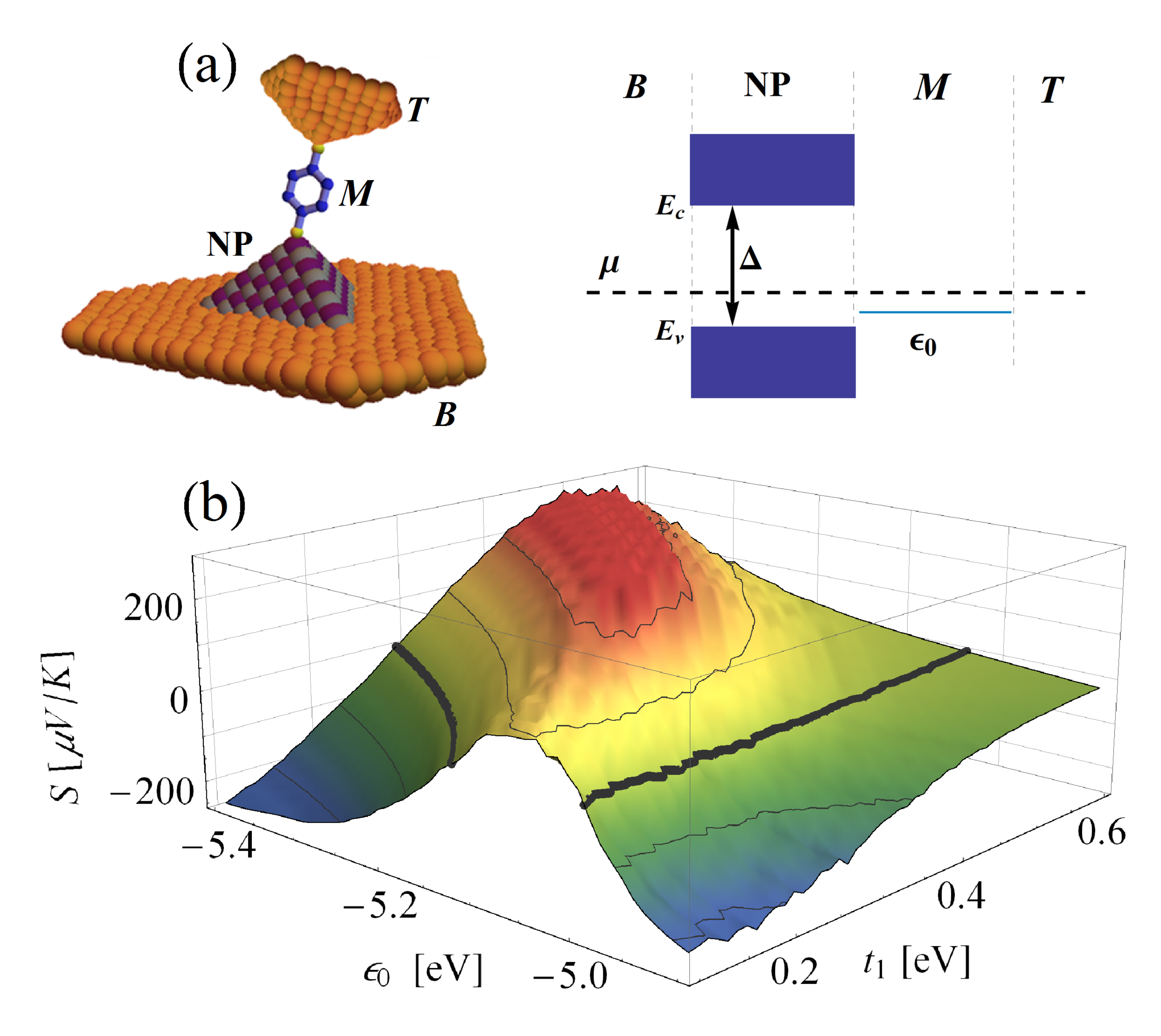}
 \caption{ (a) Schematic representation of the hybrid metal-molecule-nanoparticle-metal junctions (the choice of molecule is arbitrary), and the corresponding energy landscape.  (b) Thermopower $S$ as a function of the position of the molecular orbital $\epsilon_0$ and molecule-nanoparticle coupling $t_1$ for a hybrid molecular junction with a square-pyramid-shape (see text for numerical parameters).   }
 \label{fig1}
 \end{figure}

We start by describing the thermopower $S$ and FOM $ZT$ for a single molecule placed on a square pyramid-shaped nanoparticle \cite{Khanna2006,Wang2008}(the bottom electrode is in contact 
with the [111] plane), as shown in Fig.~\ref{fig1}(a). Since we present here a generic model for nanoparticles, we are not aiming at obtaining quantitative results describing a specific 
system. We are, nonetheless, aware that it is essential to take numerical parameters that are realistic and readily describe experimental systems. We thus choose the semiconducting band center at 
$\epsilon_c=-4.8$ eV and $\Delta =0.8$ eV, corresponding to PbSe nanoparticles, and $\mu=-5.1$ eV as the electrode chemical potential (corresponding to Au electrodes).  Other numerical parameters 
were $t=1.6$ eV , $\Gamma_B=0.05$ eV,  
$\Gamma_T=0.01$ eV (describing weakly coupled molecules, see, e.g.,\cite{Paulsson2003}). The pyramid basis contains $10 \times 10$ atoms, and our results only weakly (and quantitatively) depend on the size of the nanoparticle. 

In Fig.~\ref{fig1}(b), we show the thermopower $S$ as a function of molecular orbital energy level $\epsilon_0$ and the molecule-nanoparticle coupling $t_1$. In experiments, $\epsilon_0$ can be tuned by choice of 
molecule and by choice of the nanoparticle composition and size. The coupling $t_1$ can be additionally tuned by stretching or squeezing the molecular junction with the top electrode
 \cite{Zhou2013,Zhou2010,Diez-Perez2011}. As may be seen, $S$ can reach values as high as $\pm 200~\mu$V/K and can change sign according to the position of the molecular level with respect to the 
semiconducting band edge. 

The appearance of a thermopower maximum upon a change in $\epsilon_0$ is not surprising, since one would expect that tuning $\epsilon_0$ would lead to a near resonance in transmission (which implies 
a maximum thermopower). However, the appearance of a maximum upon a change of $t_1$ {\sl is} surprising; in a typical single-molecule junction, stretching the junction will only lead to a change in 
the molecule-electrode coupling, and will nor result in a thermopower maximum. Here, since the molecule and the NP hybridize, changing $t_1$ (experimentally - by means of pulling the junction) is 
similar to changing the molecular orbital, i.e. pulling the junction plays the role of gating. Since gating a molecular junction is a challenging task \cite{Ballmann2012,Capozzi2014,Kim2014,Prins2011}, this result puts an additional advantage on hybrid junctions.   

\begin{figure}[h!]
 \vskip 0.0truecm
 \includegraphics[width=8.5truecm]{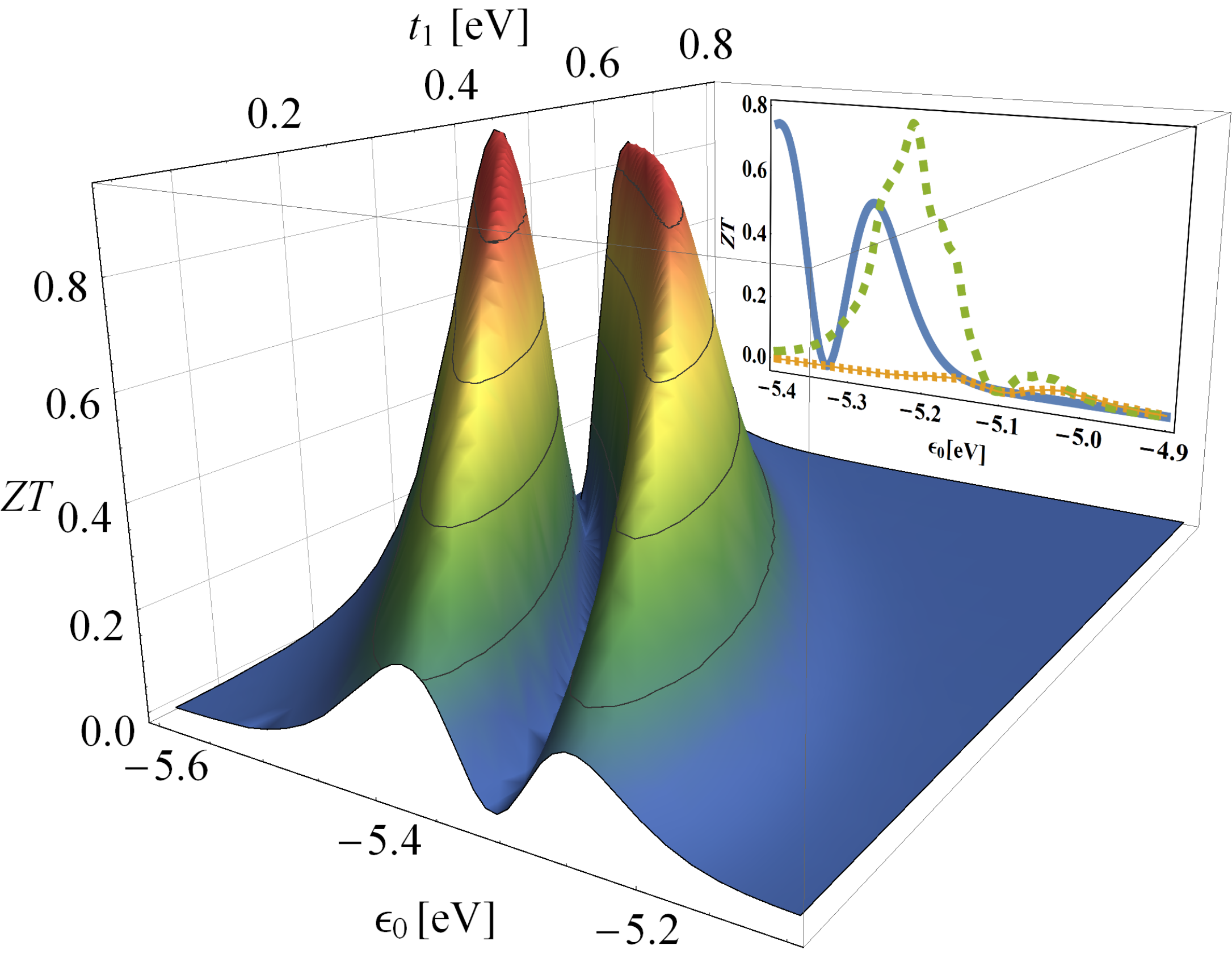}
 \caption{$ZT$ as a function of the position of the molecular orbital $\epsilon_0$ and the molecule-nanoparticle coupling $t_1$ for a hybrid molecular junction with a square-pyramid shape (see text for numerical parameters).  Inset: $ZT$ as a function of $\epsilon_0$ for a constant $t_1=0.1$ eV, for the hybrid junction with a pyramid nanoparticle (blue solid line), a pyramid nanoparticle with switched $\pm \Delta$ positions (dotted orange line, see text), and a cube-shaped nanoparticle (dashed green line).   }
 \label{fig2}
 \end{figure}

To calculate $ZT$, we add to the electronic thermal conductance a phononic term $\kappa_{phn}=50$ pW/K, a realistic value for molecular junctions \cite{Malen2010,Ong2014,Segal2003,Wang2007}. In Fig.~\ref{fig2}, we plot the 
central result of this paper, $ZT$ of the hybrid single-molecule-NP junction, as a function of $\epsilon_0$ and $t_1$. The FOM reaches $ZT>0.8$, i.e., more than three orders of magnitude larger than 
the values measured in regular molecular junctions. The realistic parameters considered here and the wide range of parameters in which $ZT$ is large implies that this regime should be accessible for experiments. 

In the inset of Fig.~\ref{fig2}, we show $ZT$ as a function of $\epsilon_0$ for a constant $t_1=0.1$ eV for three different configurations. The first (solid blue line) is the same as the setup in the
main figure. In the second setup (dotted orange line), the molecule is in contact with a pyramid nanoparticle, but the positions of the $\pm \Delta$ terms in the nanoparticle Hamiltonian of Eq.~\ref{H-NP} are 
switched, modeling a change in the atom species at the apex of the nanoparticle (for instance either Pb or Se in PbSe nanoparticles). The third setup (dashed green line) describes a molecule in contact with a cube-shaped (as opposed to a pyramid) nanoparticle. As may be seen, the contact configuration and the shape of the nanoparticle can have a strong effect on $ZT$; although values of $ZT>0.8$ can be achieved in these configurations, the optimal parameters vary between the different setups. 

\begin{figure}[h!]
 \vskip 0.0truecm
 \includegraphics[width=8.5truecm]{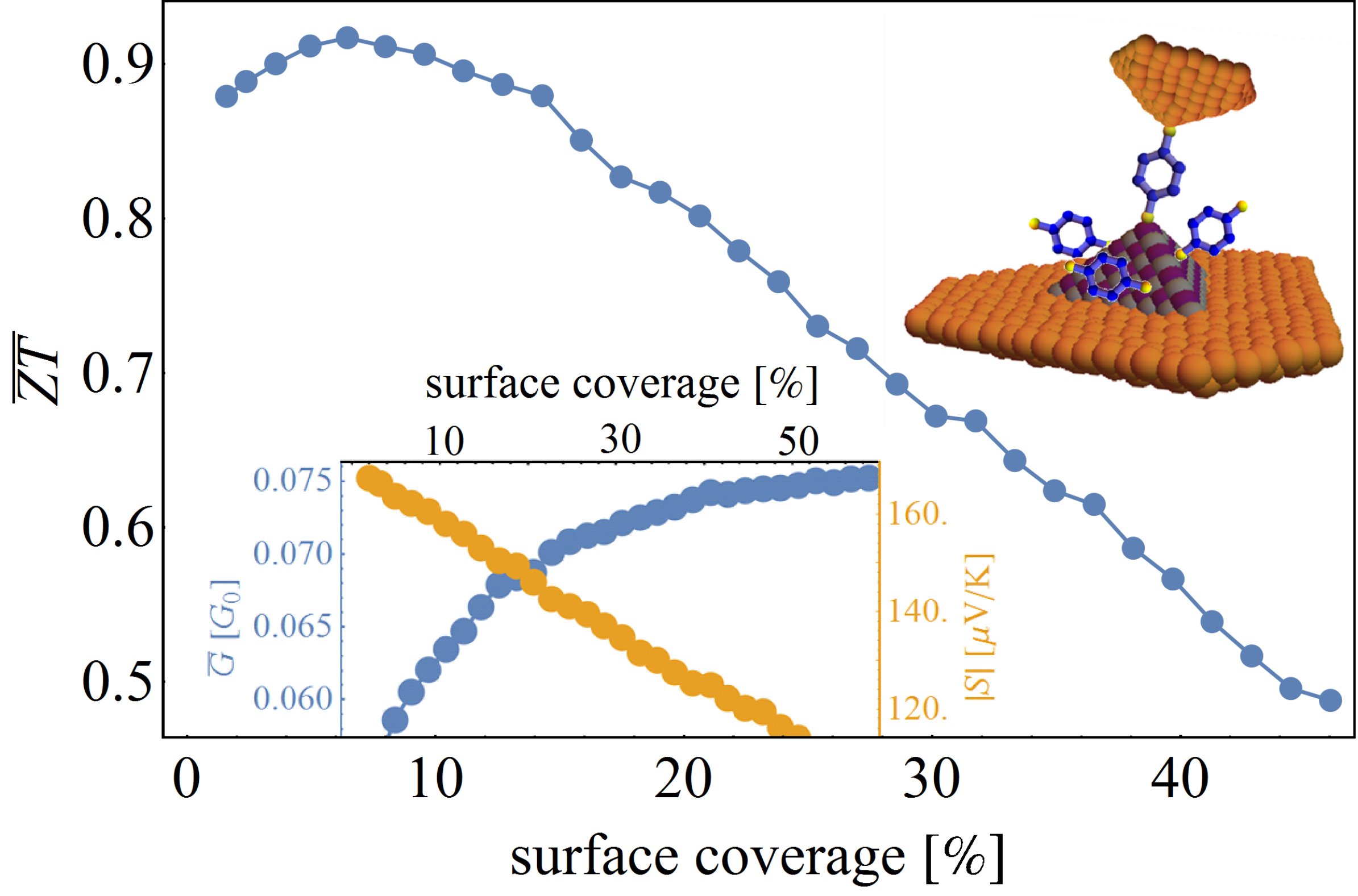}
 \caption{$ZT$ as a function of surface coverage (in percents) of molecules attached to the nanoparticle surface. Upper inset: schematic representation of the surface-attached molecules. Lower inset: average conductance (blue circles) and thermopower (orange triangles) as a function of number of surface molecules.}
 \label{fig3}
 \end{figure}

In experiments, a reasonable situation would be for additional molecules to attach to the surface of the nanoparticle. To model this scenario, we add to the Hamiltonian additional molecules, with the same 
orbital level $\epsilon_0$ and coupling $t_1$ but with a random coupling point to the surface of the nanoparticle (top inset in Fig.~\ref{fig3}). The transport properties are then averaged over $10^{4}$  
realizations of random positions. In Fig.~\ref{fig3}, $ZT$ is plotted as a function of surface coverage (in percents) of attached molecules. Surprisingly, we found a slight increase of $ZT$ for a small number ($\sim 8 \%$) of attached molecules, followed by a decrease in $ZT$ when the number of attached molecules was increased. To elucidate the origin of this result, the lower inset shows the average conductance (blue circles) and thermopower (orange triangles) as a function of surface coverage (in percents). We found that the average conductance actually increased with the number of surface molecules, but the thermopower decreased, thus eventually leading to a decrease in $ZT$.

The origin of this effect is the fact that the surface coverage induces two competing processes. On one hand, the addition of molecules bound to the surface adds conduction channels (i.e. local resonances in the transmission 
function) and therefore increases the conductance. One the other hand, the presence of disorder tends to flatten the resonances on average, and as a result the thermopower, which is proportional to the derivative of the 
transmission function \cite{Dubi2011}, and hence becomes smaller as the transmission resonance becomes wider, is reduced. This competition is reflected in the opposite trends of $G$ and $S$ in the inset of Fig.~\ref{fig3}. Since $ZT$ is the product of the conductance  which increases  and the thermopower  which decreases, it exhibits non-monotonic behavior.

At room temperature it is possible that the electron motion across the junction will not be coherent and that the 
electrons will dephase as they cross the molecule-NP interface. This may occur due to the interaction of electrons with the (soft) phonons of the NP, or due to vibrations in the position of the molecule with respect to the electrodes (but not due to interaction of electrons with the molecular vibrations, which are typically high-energy modes; such interactions were discussed in, e.g. \cite{Ren2012}). If this is the case, the formulation presented above is invalid, as it describes coherent transport. To account for decoherence, we 
note that if the electron loses its phase on the molecule-NP interface, then the NP can, in fact, be considered as a semiconducting electrode. This results in an effective SC-molecule-metal junction, which can again be described using the Landauer formula, with the appropriate choice of self-energies. 

To model the SC electrode, we recall that the imaginary part of the self-energy describes the electrode DOS. Similar to the wide-band approximation for the metallic electrode, we 
model the semiconducting electrode as constant outside the gap and zero in the gap. The imaginary part of the semiconducting self-energy can thus be written with the use of a heaviside step function
 $\Theta$ as ${\mathrm Im} \Sigma_{SC}=\frac{\Gamma_{B}}{2} \left( \Theta(|E-\epsilon_c|-\frac{\Delta}{2}) \right)$, and the real part of the self-energy determined via the Kramers-Kronig relation \cite{DiVentra2008,Mahan2000}. In the 
numerical calculations given below, the step-function discontinuity is broadened by $0.01$ eV, a value that can arise in realistic systems from lattice impurities and dislocations or thermal fluctuations.

\begin{widetext}

 \begin{figure}[h!]
 \vskip 0.0truecm
 \includegraphics[width=20truecm]{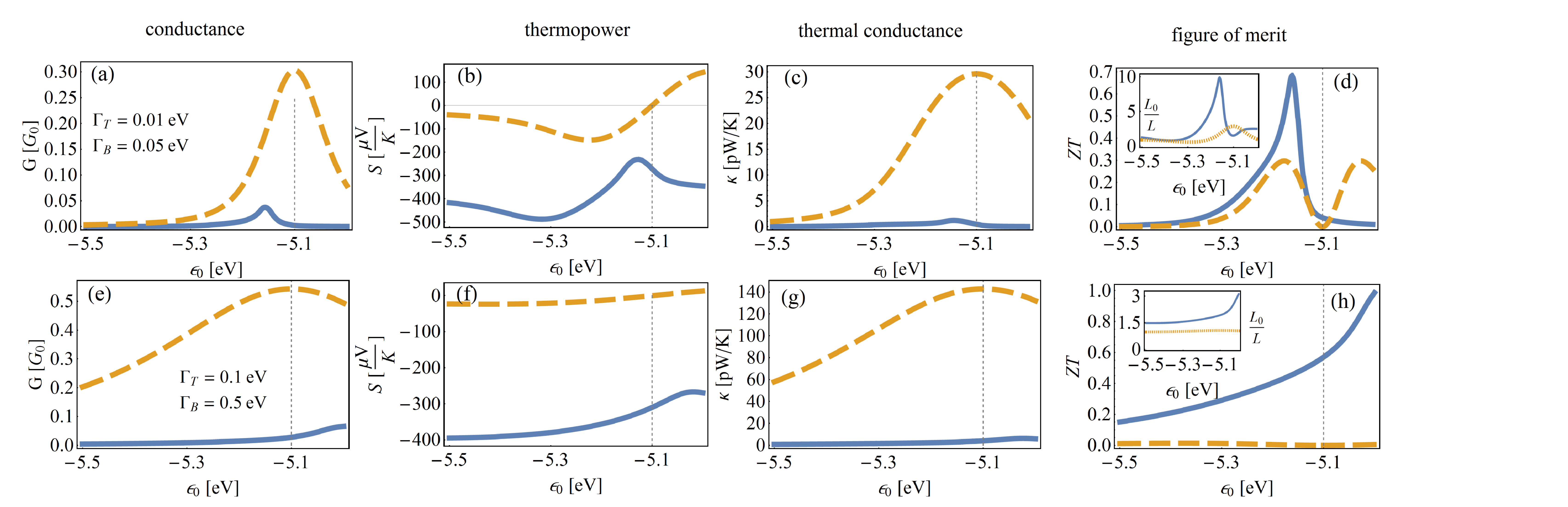}
 \caption{(a) conductance, (b) thermopower, (c) thermal conductance and (d) $ZT$ as a function of molecular orbital energy $\epsilon_0$, for a semiconductor-molecule-metal junction (solid lines) and a metal-molecule-metal junction (dashed lines), calculated for a weakly coupled molecule ($\Gamma_B=0.05$ eV, $\Gamma_T=0.01$ eV). Inset of (d): Inverse Lorenz number  as a function of $\epsilon_0$. (e-h) same as (a-d), for a strongly coupled molecule ($\Gamma_B=0.05$ eV, $\Gamma_T=0.01$ eV). }
 \label{fig4}
 \end{figure} 
\end{widetext}

In Fig.~\ref{fig4} we plot the (a) conductance, (b) thermopower, (c) thermal conductance and (d) $ZT$, for the hybrid SC-NP-molecule-metal junction (solid lines) as a function of $\epsilon_0$, with $\Gamma_B=0.05$ eV, $\Gamma_T=0.01$ eV, $\epsilon_c=-4.8$ eV, and $\Delta=0.8$ eV. Looking at $ZT$, we find that the dephasing process in the NP does not substantially reduce the maximal $ZT$ from the coherent case (the maximal value of Fig.~\ref{fig2}). For comparison, the dashed lines of Fig.~\ref{fig4}(a-d) show the 
same for $\Delta=0$, i.e. a "standard" metal-molecule-metal (M-M-M) junction. The two most striking features of the comparison between the hybrid junction and the M-M-M junction are (i) the thermopower is substantially larger for the 
hybrid junction and over a wider range of $\epsilon_0$, and (ii) $ZT$ of the hybrid junction is roughly twice as large as that of the M-M-M junction. 

The inset of Fig.~\ref{fig4}(d) shows the inverse Lorenz number $\frac{L_0}{L}$ as a function of $\epsilon_0$ for the hybrid junction (solid line) and M-M-M junction (dashed line). While a violation of the 
Wiedeman-Franz (WF) law $\frac{L_0}{L} \sim 1$ is observed for both types of junctions, the hybrid junction shows a much larger violation, which in fact determines the position 
of the maximal $ZT$, while for the M-M-M junction, the position of the maximal $ZT$ is determined by the maximum of $S$. 

Up till now, we considered a molecule which is weakly coupled to the electrodes. However, depending on the chemical moiety, the coupling between the molecule and the electrodes can be much larger (see, e.g. \cite{Bergfield2009,Bergfield2010,Paulsson2003}). It is therefore of interest to see whether the increase in $ZT$ in hybrid junctions compared to the M-M-M junctions is maintained also for strongly coupled molecules. 
In Fig.~\ref{fig4}(e-h), the same as in Fig.~\ref{fig4}(a-d) is plotted for a strongly coupled molecule, with  $\Gamma_B=0.5$ eV, $\Gamma_T=0.1$ eV. Again, the two most striking differences between the hybrid junction and the 
M-M-M junction are the thermopower and $ZT$, which are even more profound for the strongly coupled junction. Due to the large value of the coupling, the transmission resonance of the M-M-M junction is very 
broad, and therefore the thermopower is very small. That, in addition with the fact that the WF law is obeyed (shown in the inset of Fig.~\ref{fig4}(h), dashed line) results in a small $ZT \sim 10^{-2}$. In contrast, the hybrid 
junction (solid lines) shows only slight reduction of the thermopower (Fig.~\ref{fig4}(f)), because although the molecular level is broadened due to the metallic lead, the band-edge of the SC electrode still introduces a sharp 
feature to the transmission function. Along with a violation of the WF law (inset of Fig.~\ref{fig4}(h)), this leads to a relatively large $ZT\sim 1$. Surprisingly, $ZT$ for the strongly-coupled hybrid junction displays larger $ZT$ than the weakly-coupled junction, with large values of $ZT$ for $\epsilon_0$ well inside the SC band gap (as a result of the WF law violation inside the gap).

\begin{figure}[h!]
 \vskip 0.0truecm
 \includegraphics[width=8.5truecm]{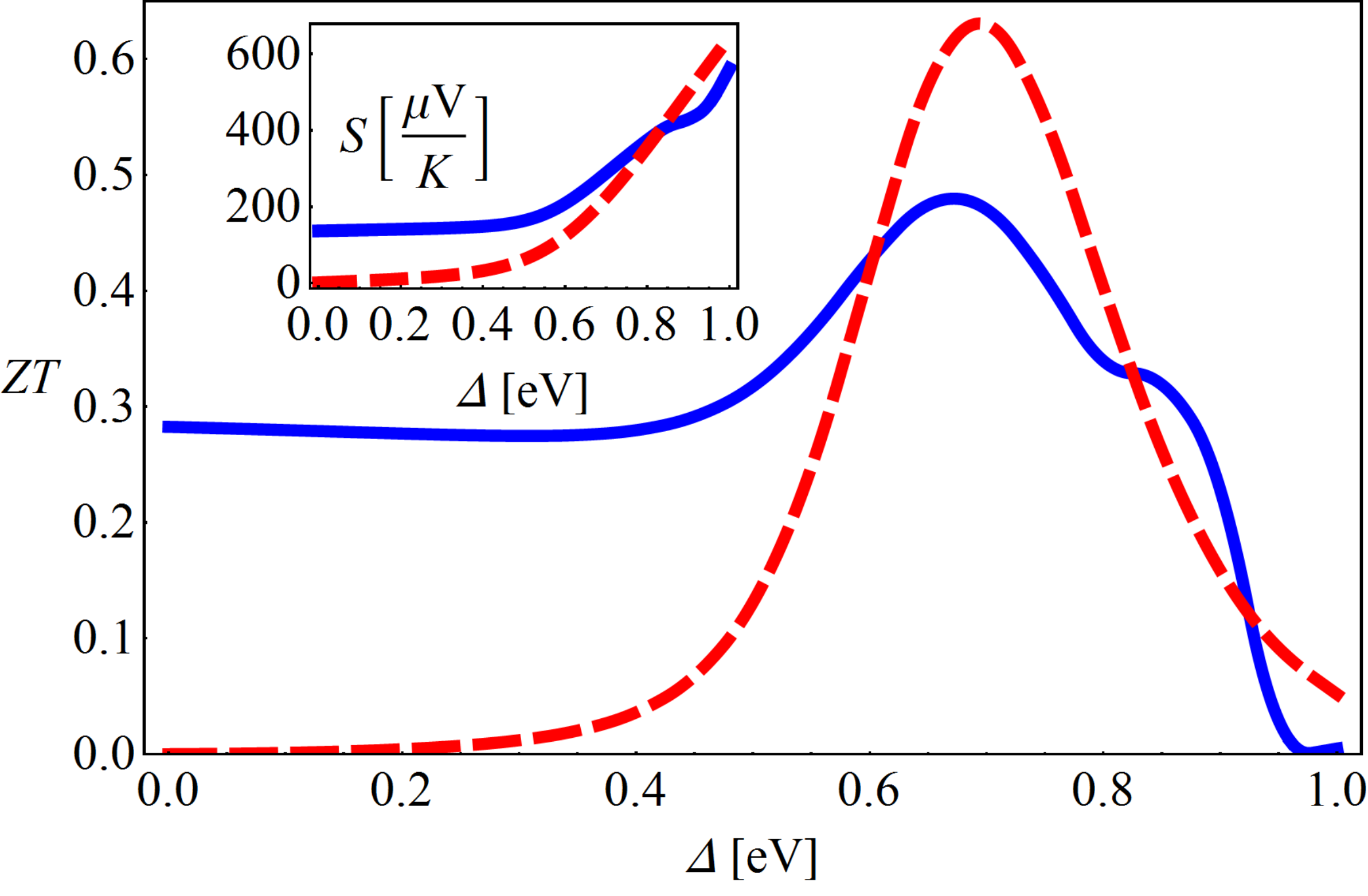}
 \caption{$ZT$ as a function of the SC NP band-gap $\Delta$ at constant molecular level $\epsilon_0=-5.2$ eV, for weakly coupled (solid blue line) and strongly coupled (dashed red line) molecule. Inset: 
$S$ as a function of $\Delta$.  }
 \label{fig5}
 \end{figure}

As was noted earlier, one of the advantages of the hybrid NP-molecule junctions is the ability to tune not the molecular orbitals, but the SC-NP band structure. With this in mind, in Fig.~\ref{fig5} we plot $ZT$ for a constant 
value of the molecular orbital $\epsilon_0=-5.2$ eV, as a function of the band-gap $\Delta$ (the band center is at $\epsilon_c=-4.8$ eV and $\Delta=0$ corresponds to a M-M-M junction, see Fig.~\ref{fig1}(a)), exploring 
the two cases of weakly coupled molecule (solid blue line) and strongly coupled molecule (dashed red line), as described above. For the weakly coupled molecule we see an increase of a factor $2$ from the M-M-M junction to the 
optimal band-gap. For the strongly coupled junction we find that $ZT$ can rise as high as $\sim 0.6$, with a four orders of magnitude increase in $ZT$ compared to the M-M-M junctions. The inset shows $S$ as 
a function of  $\Delta$, and both the weakly-coupled and strongly-coupled molecules exhibit orders-of-magnitude increase in $S$ compared to the M-M-M junctions. 

\begin{figure}[h!]
 \vskip 0.0truecm
 \includegraphics[width=8.5truecm]{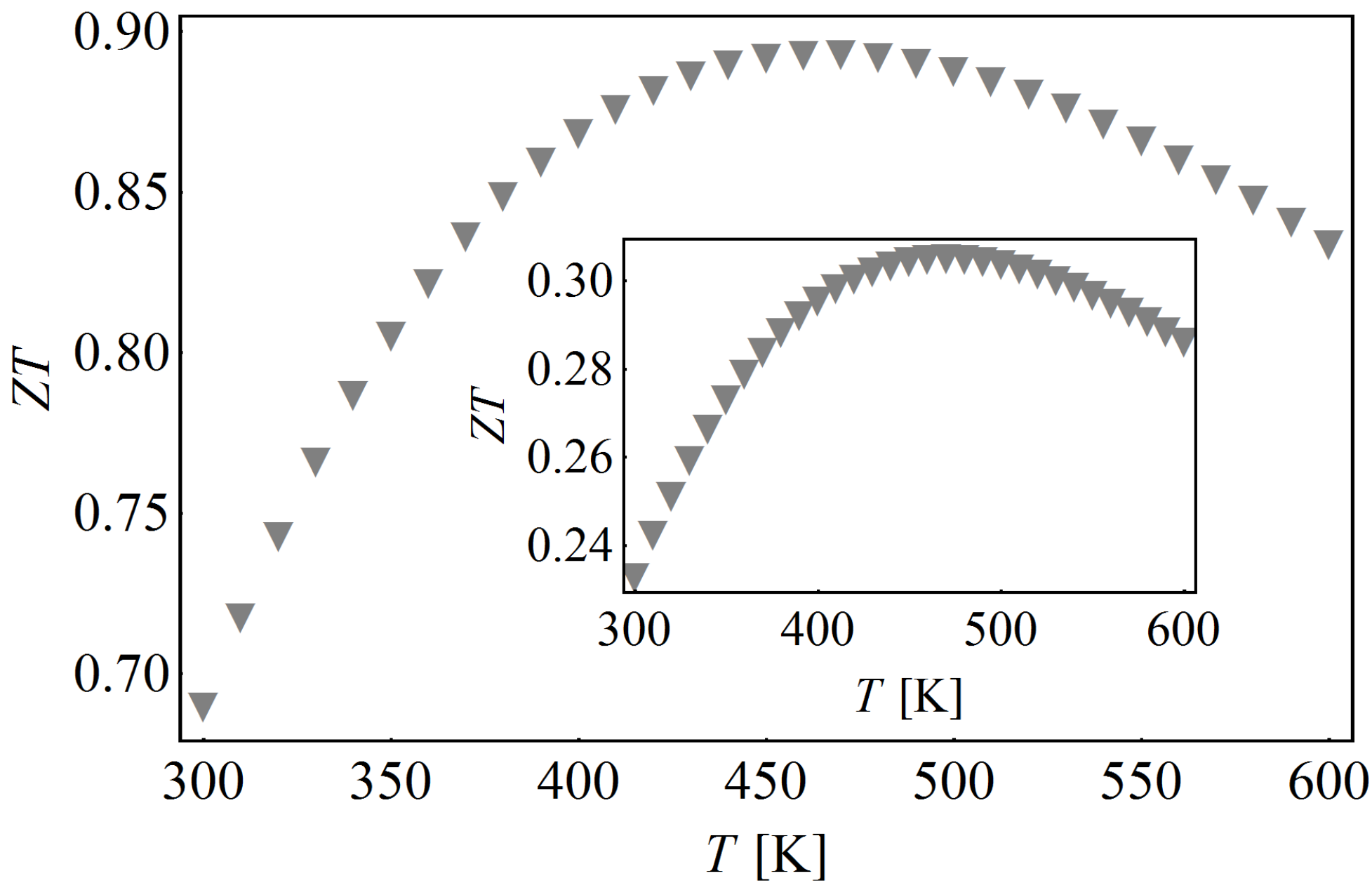}
 \caption{Temperature dependence of $ZT$ for $\epsilon_0=-5.16$ eV. Inset: same, but with $\kappa_{phn}=150$ pW/K. }
 \label{fig6}
 \end{figure}

Finally, in Fig.~\ref{fig6}, the temperature dependence of $ZT$ is examined for the weakly-coupled hybrid molecular junction, evaluated for $\epsilon_0=-5.2$ eV (corresponding to the maximum in $ZT$ from Fig.~\ref{fig4}(d)) (the 
rest of the parameters are the same as in Fig.~\ref{fig4}(a-d)). $ZT$ is found to  increases with temperature, 
exhibiting a maximum of $ZT \sim 2$ at $T\sim 450$ K, followed by a moderate decrease. This finding again implies that 
relatively large values of $ZT$ persist in a broad range of parameters. In fact, in the inset of Fig.~\ref{fig5}, we plot $ZT$ calculated with an over-evaluated value of the phonon thermal conductance $\kappa_{phn}=150$ pW/K, and we find that $ZT$ reaches values of  $ZT \sim 0.3$, which is still several orders of magnitude larger than the observed values. 

\section{Summary and conclusions}
In summary, we presented calculations of thermopower and thermoelectric FOM for a hybrid metal-single molecule- semiconducting nanoparticle-metal molecular junction. The presence of the SC-NP 
and the position of the molecular orbital close to the semiconducting band edge enhanced the particle-hole asymmetry required for efficient thermoelectric conversion. The resulting values of 
thermopower and $ZT$ were much larger than those measured in experiments, and reach values as high as $S\sim 500~\mu$V/K and $ZT \sim 1$ (an improvement of four orders of magnitude improvement over measured molecular junctions). 
We showed that this enhanced thermoelectric performance persists over a wide range of parameters and is robust against disorder in the form of surface-attached molecules. We showed that decoherence at the molecule-nanoparticle 
boundary is not detrimental to the thermoelectric performance, and that large values of $ZT$ persist up to high temperatures. Comparing hybrid SC-NP-molecule junctions to the more "standard" 
metal-molecule-metal junction, we found that for weakly-coupled molecules there is a factor $\sim 2$ increase in $ZT$ and a factor $\sim 4$ in the thermopower. For strongly coupled molecules, the advantage of hybrid NP-molecule junctions is even more profound, with $2-4$ {\sl orders of magnitude} increase in $ZT$ and thermopower n hybrid junctions.

The model we presented here is a generic model, not aimed at any specific system. Nevertheless, our numerical parameters were taken from experimentally observed value, including the phononic 
contribution to the thermal conductance. This, along with the fact that enhanced thermoelectric performance was found for a wide range of molecular parameters and was robust against disorder, 
decoherence and high temperatures, strongly suggest that high values of thermopower can be reached in future experiments on molecule-nanoparticle junctions, which are promising candidates for 
nano-scale thermoelectric conversion. 

\section{Acknowledgments} 
The authors acknowledge funding from the BGU-UM joint research initiative (Adelis Foundation).


%

\end{document}